# Sodium and Potassium Linewidths as an Atmospheric Escape Diagnostic at Mercury

P. Lierle[1] and C. Schmidt[1]

[1] Center for Space Physics, Boston University, Boston, MA, USA

Corresponding author: Patrick Lierle (plierle@bu.edu)

**Key Points:**

- Mercury's potassium tail was observed for the first time. At a downtail distance of 5.8 $R_M$ the Na/K ratio is ~95.
- Linewidths broaden with downtail distance. By 5 $R_M$, thermal widths of sodium increase from 1200 to 7500 K and potassium from 750 to 8500 K.
- Line broadening is attributed to gravitational filtering of the velocity distribution function, a novel diagnostic for atmospheric escape.

**Abstract**

The spatial distribution and linewidth of Mercury's sodium and potassium exosphere were observed using a combination of long-slit and high-resolution point spectroscopy. Effective temperatures were estimated from emission line profiles by forward modeling their Doppler broadening. These serve as an energy metric for collisionless gas that is inherently nonthermal. The Na gas at low and mid-latitudes ranges from 1200-1300 K along the noon meridian, in agreement with MESSENGER scale heights, increasing by ~200 K at the poles and terminator. This increase is attributed to the loss of low energy atoms to the surface during photon-driven transport antisunward. An escaping potassium tail was measured for the first time, observed to a distance of 10.4 $R_M$ with Na/K ~95 at 5.8 $R_M$. Emission linewidths increase sharply between the dayside and escaping tail, with Na growing from about 1200 to 7500 K, and K from 750 to 8500 K by the time the gas reaches 4.3 $R_M$ downtail. Na D line profiles down the exotail also evolve from Gaussian to boxcar in shape. Both characteristics are interpreted as filtering of the nascent velocity distribution function, wherein low energy atoms on gravitationally bound trajectories are removed from the gas population, while high energy escaping atoms are retained. Na linewidths become invariant past 3.5 $R_M$, placing this altitude as the ballistic apex of bound trajectories. In this way, Mercury's emissions prototype a novel technique towards a broader understanding of atmospheric escape, using emission line morphology to probe the transition between bound and escaping gas.

**Plain Language Summary**

The small amount of gas comprising Mercury's atmosphere presents a unique observational environment due to its proximity to the Sun. Pressure from solar photons forces this atmosphere into a comet-like tail. Here, the spatial and energy distribution of sodium and potassium gas were observed with ground-based telescopes, both near the surface and in the escaping tail. The energy distribution of the gas is represented by effective temperature, a quantity derived from the phenomenon that more energetic gases produce wider emission lines via the Doppler effect. On Mercury's disk, sodium effective temperatures are cold, in agreement with past measurements from the MESSENGER orbiter. Effective temperatures increase dramatically down both the sodium and the potassium tails, the latter detected for the first time. This increase is attributed not to heating but rather to the gravitational loss of low energy atoms. As measurements are taken farther and farther from the planet, more and more gravitationally bound atoms will be absent from the observed energy distribution. With only the hottest atoms reaching the highest altitudes, emission lines appear hot, despite no heating. These measurements of false broadening in emission lines represent a technique that has not been previously applied in studies of atmospheric escape.

## 1. Background

In the absence of inter-particle collisions, the energy distribution of gases in Mercury's surface-bound exosphere does not necessarily conform to the canonical Maxwell-Boltzmann distribution. Still, this energy distribution is a fundamental characteristic of the physical processes that supply and remove the exosphere, and so effective temperature, if this gas were approximated as thermal, is a useful quantity. Moreover, the two conventional methods to evaluate gas energy by remote sensing are both rooted in the assumption that the gas population is thermally distributed. These are 1) the atmospheric scale height, for which Chamberlain (1963) derived corrections to the barometric law appropriate for planetary coronae, and 2) the Doppler broadening of emission lines, for which derivations appear widely in spectroscopy textbooks (e.g., Emerson, 1996).

Effective temperatures obtained by these two approaches do not always agree, but calcium in Mercury's exosphere offers an example of how this disagreement itself can be insightful, and not merely discrepancy that points to erroneous analysis. Ground-based high-resolution spectroscopy first detected the neutral calcium emission line at 4227Å with Doppler broadened effective temperatures ranging from 12,000 to 20,000 K (Bida et al., 2000; Killen et al., 2005). The atmospheric scale height of this same emission line obtained from the Ultraviolet and Visible Spectrometer (UVVS; McClintock & Lankton, 2007) instrument on the Mercury Surface, Space Environment, Geochemistry, and Ranging (MESSENGER) orbiter was much hotter, however, in the 50,000 to 70,000 K range (Burger et al., 2014). The tension between these discrepant quantities is not yet fully understood, but a working theory can explain the difference qualitatively: a source of Ca that is distributed with altitude, as opposed to supply purely from the surface.

Vapor created by micrometeoroid impacts is locally collisional, enabling chemical reactions to create gas phase products like CaO and CaS (Killen et al., 2016). Sunlight is the only viable means for such molecules to be broken apart into atomic fragments, which are imparted significant kinetic energy when the ionic bond in broken. UVVS measurements of the ejecta plumes from individual cm-sized impactors support this theory on the chemical pathway to atomic Ca. Cassidy et al. (2021) showed that an impact near the anti-solar point produces Na and Mg atoms with effective temperatures 10,000 to 15,000 K, and in stoichiometric ratio with local regolith composition. However atomic Ca, which the UVVS instrument was most sensitive to, was absent from the ejecta plume, explained by calcium remaining in a bound molecular form while devoid of sunlight on the planet's nightside. In this way, it is thought that the hotter atmospheric scale height of calcium reflects a two-part process, wherein Ca-bearing molecules are released from the surface and then photodissociated at altitude. This interpretation needs further testing with physical chemistry models, but it is qualitatively consistent with the observations: Ca gas that is hot in its kinetic linewidth, but hotter still in its altitude distribution.

Unlike calcium, the scale height and linewidth-based estimates of effective temperature agree quite well for Mercury's sodium gas. Accounting for hyperfine structure, linewidth analysis by Killen et al. (1999) focused exclusively on the $D_2$ line at 5890Å, and determined temperatures between 750 K and 1500 K, with the equatorial regions being hottest. With an added term to account for compression by solar radiation pressure, Chamberlain models fit to the altitude profiles of sodium emission in low latitude regions gave effective temperatures ranging 1100 to 1250 K (Cassidy et al., 2015). Lierle et al. (2022) reported Na $D_1$ and $D_2$ Doppler widths in this same range, and a K $D_1$ Doppler width of ~700 K. Both effective temperatures are consistent with time-of-flight laboratory experiments of alkali desorption induced by electron transfer in ionically bound atoms, and photon-stimulated desorption by blue and near-UV sunlight is the most efficient such mechanism to liberate atoms from the topmost regolith into the exosphere (Yakshinskiy & Madey, 2003). Unambiguous measurements of the K altitude profile have not yet been made since neither the MESSENGER UVVS nor BepiColombo Probing of Hermean Exosphere by Ultraviolet Spectroscopy (PHEBUS; Quémerais et al., 2020) spectrographs can measure far red wavelengths. While PHEBUS has a broadband channel that passes the faint blue potassium lines, emissions from other species are captured and their signatures cannot be disentangled.

At ~1000 km altitude and above, the slope of the sodium altitude profile abruptly transitions to a hot secondary component with effective temperature 5,000 to 20,000 K (Cassidy et al., 2015). This trace hot population has not yet been isolated in analyses of high-resolution emission line profiles. It remains debated if it is sourced via ion sputtering or micrometeoroid vaporization, though modeling favors the latter (Killen et al., 2022; Suzuki et al., 2020). Ion sputtering is highly variable depending on solar conditions, in contrast to all other release processes, and ground-based observers have attributed dynamics in the sodium exosphere to solar wind coupling with the surface (Mangano et al., 2015; Orsini et al., 2018, 2024). Ion sputtering is also expected to occur at high latitudes, where plasma is channeled to the surface through the magnetic cusps. Some observations find the Na exosphere is consistently brightest around 50–60° latitudes, near the dayside open-closed field line boundary (Schmidt et al., 2020). However, Na scale heights above both poles during MESSENGER's M3 flyby had e-folding scale heights of ~200 km or 1500 K (Vervack et al., 2010), indicating that the bulk photo-desorbed population remains predominant, at least at low altitudes. While the secondary hot sodium population will mostly escape down Mercury's comet like tail, even here it appears so rarified as to evade detection; aside from the occasional impactor plume, the Na tail exhibits characteristics entirely consistent with photo-desorption as the sole supply, both in terms of its spatial width (Schmidt et al., 2012) and its seasonal modulation (Cassidy et al., 2015).

The observations herein were intended to spectrally resolve emission lines around Mercury's disk, particularly any signatures of the hot secondary sodium population, and to potentially detect and characterize Mercury's potassium tail, which enables interesting comparisons with sodium. Section 2 introduces the observations in the context of Mercury's seasonal changes in scattering and radiation pressure. Section 3 presents long-slit low-resolution measurements characterizing the spatial structure of the Na and K exotails during peak atmospheric escape. Section 4 describes the method for obtaining effective temperatures from spectrally resolved emission line profiles. Section 5 describes sodium effective temperatures over Mercury's disk. Section 6 describes sodium and potassium line profiles in the planet's exotail. These observational results come from independent instruments, but are interrelated, with their combined interpretation described in Section 7 and summarized in Section 8.

## 2. Observations in the Context of Mercury's Seasons

Four observational campaigns were conducted to study sodium and potassium emissions at Mercury: one with a long-slit medium-resolution spectrograph to characterize the spatial distribution in the tail, and three with precision radial velocity spectrometers for point measurements of the emission line profiles at various locations in the dayside and tail. The first campaign utilized the DeVeny long-slit spectrograph at the 4.3 m Lowell Discovery Telescope (LDT) to measure cross-tail profiles of sodium and potassium in 2017, thereby informing our pointing strategy in the later three campaigns. The Keck Planet Finder (KPF) at the 10 m Keck I Telescope observed sodium linewidths at regions around Mercury's disk in 2024, while the Extreme Precision Spectrometer (EXPRES) at LDT observed sodium and potassium linewidths down the tail in 2022 and 2023. Both KPF and EXPRES feature an atmospheric dispersion correction system, which given the high airmass is critical to ensure that Na and K measurements correspond to the same spatial location.

Table 1: Observational Parameters

| Date (UT) | Instrument / Telescope | Target | Phase (°) | TAA (°) |
|---|---|---|---|---|
| 2017 Apr 1 | DeVeny / LDT | Downtail Distribution | 89 | 41 |
| 2022 Apr 25 – 28 | EXPRES / LDT | Downtail Linewidths | 88 – 99 | 65 – 80 |
| 2023 Apr 9 – 12 | EXPRES / LDT | Downtail Linewidths | 89 – 103 | 50 – 66 |
| 2024 Mar 21 – 25 | KPF / Keck 1 | On-disk Linewidths | 80 – 95 | 22 – 38 |

Each of these four campaigns occurred at different locations in Mercury's orbit, with observational parameters described in Table 1. Their associated True Anomaly Angles (TAAs) from perihelion are marked in Figure 1, showing photon scattering and radiation acceleration for Na and K at these observation times in the context of seasonal modulation. Brighter emission and stronger atmospheric escape occur when there is positive feedback in radiation acceleration as atoms Doppler shift away from the Sun (Potter et al., 2007), peaking at TAAs of 48° for K and 64° for Na. Our campaigns targeted this season exclusively.

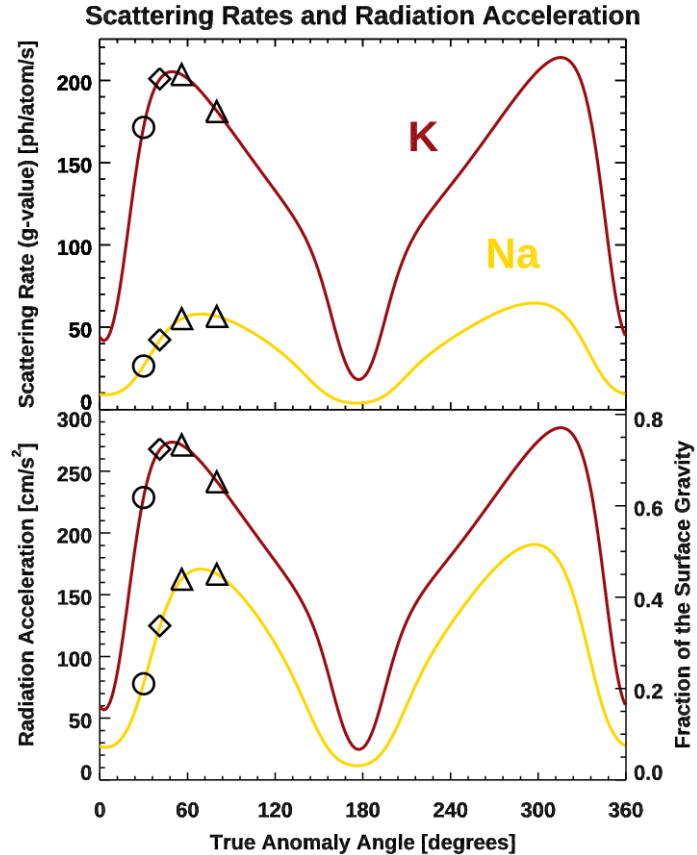

*Figure 1.* *The photon scattering rate of the combined D line doublets and corresponding radiation acceleration on Na and K atoms as a function of true anomaly angle. Seasonal modulations are slightly asymmetric and not centered at 0° and 180° TAA due to solar gravitational redshift. The four observing campaigns described herein are marked, the circle denoting KPF, the diamond, DeVeny, and the triangles, EXPRES.*

Note that seasonal modulation is not centered at 180° TAA (aphelion), and shows a slight asymmetry where scattering in the inbound leg of Mercury's orbit (>180° TAA) exceeds that in the outbound orbital leg. This is a consequence of energy loss via general relativity as solar photons escape the Sun's gravitational well. Gravitational redshift offsets the Fraunhofer absorption lines by 0.633 km/s relative to the rest of the solar system, and so solar Na and K are Doppler shifted relative to the exosphere, even when Mercury has zero heliocentric velocity. Analysis of data obtained during Mercury's transits of the solar disk brought this effect to our attention (Schmidt, 2022), and it has not been described previously; theoretical seasonal modulations in photon scattering that are symmetric with respect to the planet-Sun geometry broadly appear in past literature, but the solar gravitational redshift breaks this symmetry.

### 3. Long Slit Observations of Mercury's Exotail

Measurements of Mercury's escaping tail, hereafter referred to as the exotail, can only be made during brief twilight periods that optimize two geometric conditions simultaneously: the planet's season should be just after its perihelion, when atmospheric escape is maximized, and it should be near its quadrature with the Earth and Sun. Quadrature apparitions give solar elongation angles

>20° with the tail oriented in the plane of the sky so that its projection along the line of sight is minimal. This coincides with a high Earth-Mercury radial velocity, which helps to separate telluric and Hermean emissions and shifts potassium $D_2$ emission away from an overlapping telluric $O_2$ absorption band. Such conditions were met on 1 April 2017, with an observer phase angle of 89° at 41° TAA. DeVeny was configured with a 1200 line/mm grating and a 0.9″ wide slit, giving a resolving power of R~5,000. The slit length was oriented along Mercury's north-south axis and offset by roughly 5 and 10 Mercury radii ($R_M$) downtail, as shown in Figure 2. Integration times ranged from 300-500 seconds.

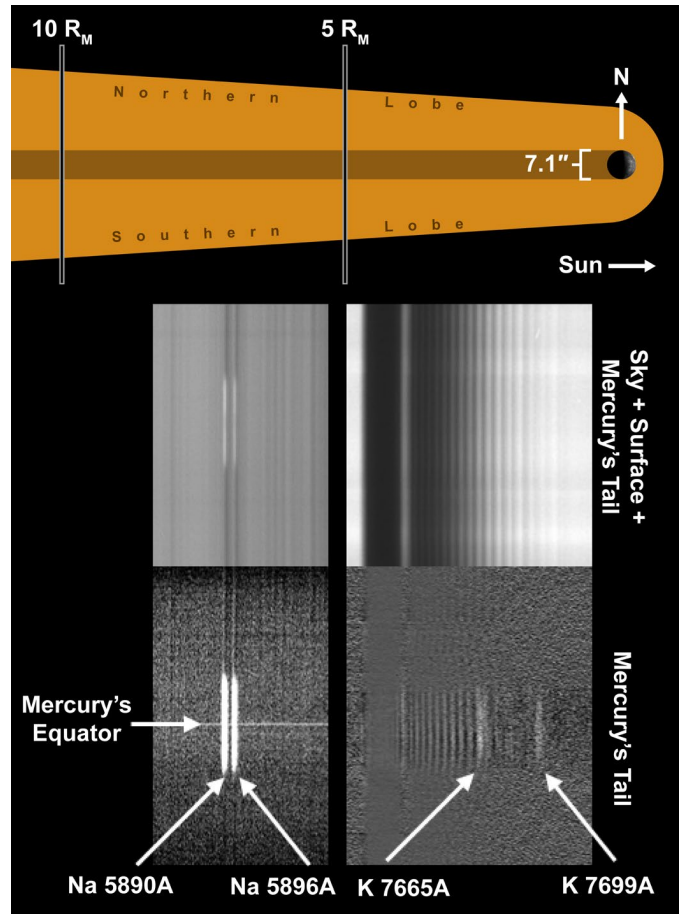

*Figure 2.* Long-slit spectra of Mercury's tail consist of exospheric emissions, twilight sky, and off-axis scattered sunlight from Mercury's bright dayside surface. The latter two must be removed to isolate tail emissions. The top panels show raw 2D spectra near 5 $R_M$ downtail before this subtraction, and the bottom panels, after. The pixel row at Mercury's equator is highlighted as a fiducial in the lower left panel.

Exospheric emissions were isolated by subtracting the interpolated sky background and a scaled spectrum of Mercury's dayside. North-south cross sections of the Na and K tails are given in Figure 3. Absolute flux calibration is obtained via photometric modelling from the MESSENGER Mercury Dual Imaging System (MDIS) reflectance (Lierle et al., 2022, 2023; Schmidt et al., 2020), but as the DeVeny guide camera saturates at low airmass, an extinction model must be applied, which is dubious at such airmass. Without a direct fiducial for atmospheric haze and only model extrapolation, uncertainty in absolute brightness and column density cannot be quantified. However, our objective is spatial information and comparisons between Na and K, which require relative, not absolute, flux calibration.

These measurements constitute the first detection of Mercury's potassium tail. Telluric $O_2$ absorption severely limits results at low spectral resolution and strong artefacts remain, but both lines of the potassium D doublet are present in the expected ratio. The brightness of the tail dips significantly in the absence of photon scattering within the planet's shadow where column density cannot be quantified. In sunlight at just over 5 $R_M$ downtail, the mean Na/K ratio is 95. This is consistent with the 70–130 ratio on Mercury's disk (Leblanc et al., 2022; Lierle et al., 2022), but only out of coincidence. Potassium gas has a lower effective temperature, making it more gravitationally bound than sodium, but also experiences greater radiation acceleration per Figure 1, which preferentially strips it. Evidently, these two approximately balance so that the Na/K ratio in the escaping tail is similar to that in the bound exosphere near the planet, at least during this season.

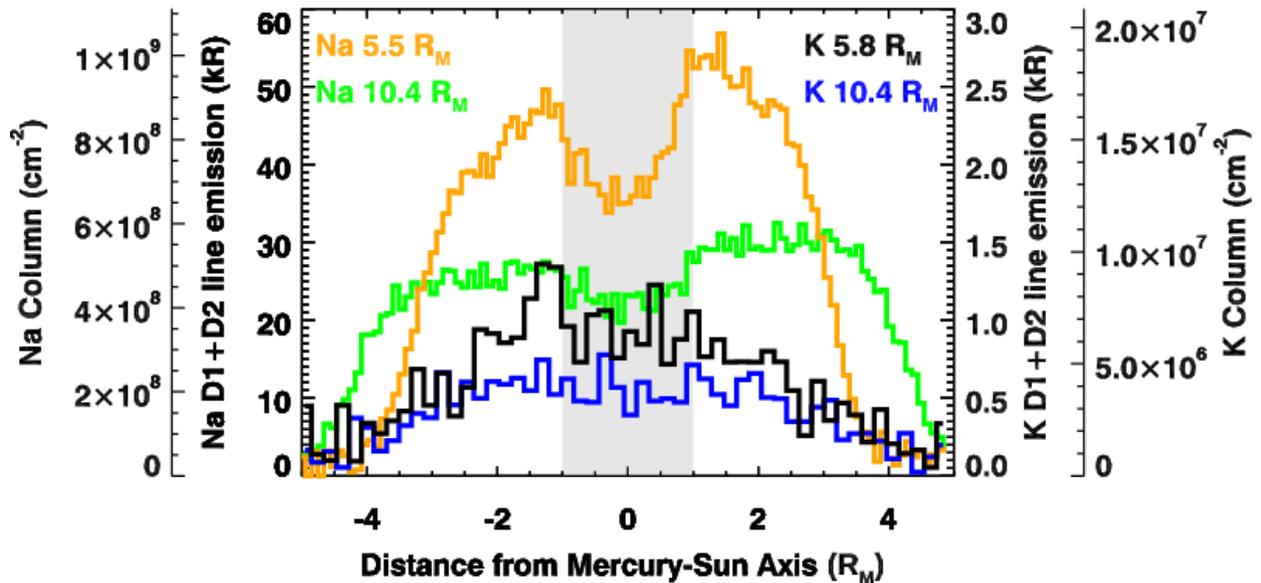

*Figure 3. North-south cross-sections of Na and K emission in the tail from LDT on 1 April 2017. Mercury's true anomaly angle is 41° and the observer phase angle was 89°. Column density cannot be quantified within the planet's shadow (shaded).*

Higher sodium abundance in the northern lobe of the exotail, seen in Figure 3, is a persistent asymmetry as shown by Potter & Killen (2008), and modeling by Schmidt (2013) demonstrated it is a consequence of enhanced sodium sources at Mercury's southern (opposite) hemisphere. The width of the potassium tail appears similar to that of sodium, with peak emission near the boundary line of the planet's shadow. While the southern lobe of the K tail appears brighter at 5.8 $R_M$ downtail, poor signal with strong $O_2$ artefacts precludes an interpretation that this is indeed physical. Overall, observational attempts with this facility could only rarely detect Mercury's escaping potassium, while all showed a north/south sodium ratio of unity or greater in the tail. From these measurements, it is clear that long-slit high-resolution spectroscopy is a superior approach to disentangle faint K emission from telluric artefacts, and rather than risk overinterpretation, further characterization of the potassium exotail's distribution is deferred until such observations become available. The sections to follow focus on high resolution point measurements.

## 4. Forward Modeling Sodium and Potassium D Line Profiles

The D transitions in sodium and potassium occur between the ground and the first excited state in their respective atoms. The first excited state for both these species is split by spin-orbit coupling to produce two fine structure transitions, $D_1$ and $D_2$. For sodium, $\lambda_{D1} = 5896$ Å, $\lambda_{D2} = 5890$ Å, and for potassium $\lambda_{D1} = 7699$ Å, $\lambda_{D2} = 7665$ Å. Each of these transitions is further divided into several hyperfine transitions by anomalous Zeeman splitting. For sodium, four hyperfine components blend to make up the $D_1$ line, and six make up $D_2$ (Brown & Yung, 1976). Energy levels are much closer together in both potassium excited states (Falke et al., 2006), and so only two hyperfine components from Morton (2003) were necessary to account for the splitting of the ground state.

The absorption coefficient for each hyperfine component is formally the convolution of its Lorentzian natural absorption profile with a Gaussian Doppler profile, but the former can be neglected as its width is < 1 mK. The Gaussian full width at half maximum (FWHM) for a transition with line center $\lambda_0$ is given by:

$$\Delta\lambda_D = \sqrt{\frac{8kT\ln(2)}{m}}\frac{\lambda_0}{c} \qquad (1)$$

where $k$ is Boltzmann's constant, $m$ is atomic mass, $c$ is the speed of light, and $T$ has Kelvin units. This Doppler width can be combined with the velocity distribution to yield a photon absorption coefficient for each hyperfine component:

$$\alpha_\lambda = \left(\frac{\pi e^2}{m_e c}\right) f \frac{2(\ln 2)^{\frac{1}{2}}}{\sqrt{\pi}\Delta\lambda_D} \exp\left[-\ln 2 \left(\frac{2(\lambda - \lambda_0)}{\Delta\lambda_D}\right)^2\right] \qquad (2)$$

where $e$ and $m_e$ are the charge and mass of the electron and $f$ is the transition's unitless oscillator strength. Note that Equation (2) is equivalent to Brown & Yung (1976) Equation (15) except for a factor of $2^2$ inside of the exponential term, which they had omitted.

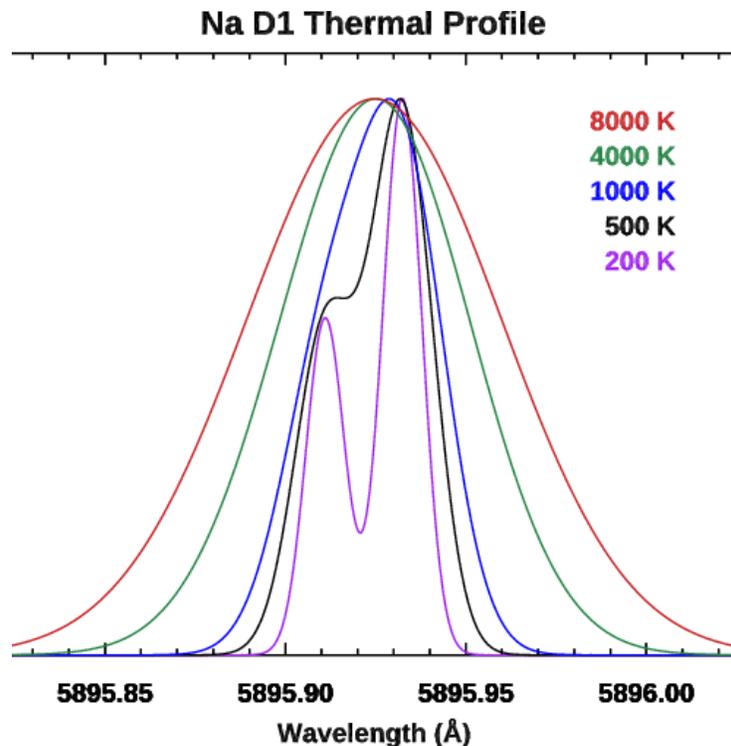

*Figure 4.* Doppler-broadened thermal line absorption profiles for the sodium $D_1$ transition. Hyperfine structure arising from the splitting of the ground state is apparent below 1000 K.

Summing over each hyperfine absorption component produces an optically thin thermal line profile, as shown for sodium $D_1$ at several temperatures in Figure 4. Asymmetric hyperfine structure is evident below 1000 K. Above 1000 K, the line profile appears roughly Gaussian in shape, but it is far wider than a Doppler-broadened single component emission line at the same temperature due to the splitting of the ground state. This line profile is what a spectrometer of infinite resolution would measure. To forward model a given measurement, it must be convolved with the instrumental line spread function (LSF) intrinsic to each instrument, which can be characterized using a spectrum of a cold spectral lamp or laser frequency comb. In this analysis, Gaussian profiles are fit to all Th-Ar hollow cathode emission lines within ~3Å of the wavelength of interest, and the minimum linewidth is taken to be the local instrumental resolution. Thermal profiles as in Fig. 4 are convolved with a Gaussian of this instrumental width and least-squares fit to the measured exospheric line profile, keeping fine wavelength shift, amplitude, and temperature as free parameters. This generates a best-fit effective temperature, with associated uncertainty propagated from Poisson statistics on the measured continuum and the uncertainty in the instrumental line fit.

## 5. Sodium Doppler Broadening over Mercury's Disk

Mercury was observed 21-25 March 2024 at the 10 m Keck 1 telescope at Mauna Kea, HI using the newly commissioned Keck Planet Finder. Observations were scheduled near greatest solar elongation to minimize the projection of the bulk motion of atoms down the tail along the line-of-sight. Mercury's phase angle varies from 80-95°, and its TAA from 22-38°. The 1.1″ diameter (730 km) KPF fiber was pointed at seven regions across the disk: the subsolar point, northern and southern magnetic cusps, north and south poles, terminator, and nightside limb. Additionally, measurements sampled 1 $R_M$ above the north and south poles, and 2 and 4 $R_M$ down the northern lobe of the tail. Integration times ranged from 45-100s on the illuminated portion of the disk, and from 200-400 s above the poles, on the nightside, and downtail. Figure 5 shows a 100 s KPF integration at Mercury's south pole, demonstrating the unprecedented signal to noise that can be obtained with Keck's large aperture and 32-bit detector, in which nearly a million photo-electrons from Mercury's exosphere are recorded per high resolution spectral bin.

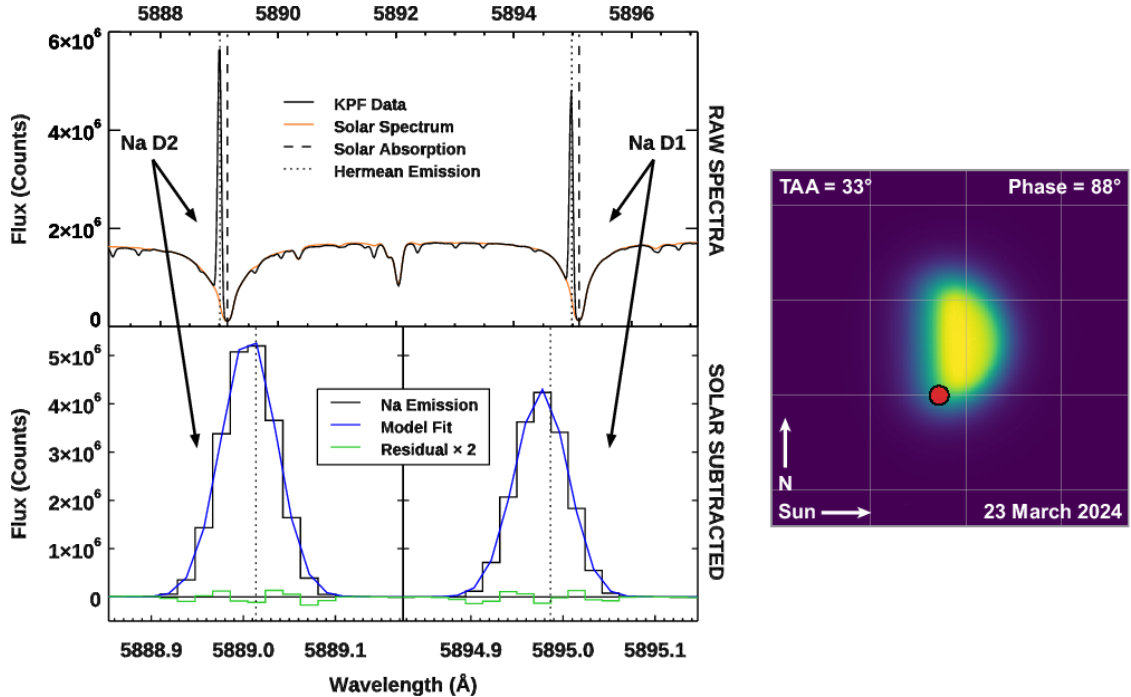

*Figure 5.* Left Top: A 100 s Keck 1/KPF spectrum of Mercury's South pole acquired at the Keck 1 telescope on 23 March 2024. Left Bottom: Extracted exospheric sodium $D_1$ and $D_2$ emissions overlaid with best-fit thermal models (blue) and data-model residuals multiplied by 2 for visibility (green). Right: Guider image displaying Mercury's dayside emission with 1.1" diameter KPF fiber position in red.

KPF data are processed through an initial pipeline that performs standard calibrations and extracts 1D spectra and wavelength solutions from the 2D green and red CCD images (Gibson et al., 2020). The KPF science fiber is sliced into three waveguides, recorded individually to improve spectral resolution. After initial processing, the 1D spectrum of each is summed. This resultant spectrum again consists of emission from sunlight reflected off Mercury's disk, exospheric emission, and sky background, which is negligible for on-disk observations. To remove the reflected sunlight component, the TSIS-1 Hybrid Solar Reference Spectrum (Coddington et al., 2021) is shifted to match the two-way Sun-Mercury-Earth Doppler shift. The solar reference spectrum is then least-squares fit to the KPF continuum with free parameters being fine wavelength shift, an additive offset, a multiplicative factor, and a Gaussian convolution kernel to match the KPF resolving power. Once fit, the reflected solar component is subtracted, leaving only exospheric emission.

Figure 5 shows this process on sodium D for a sample 100 s spectrum near Mercury's south pole with a best-fit effective temperature of 1394 ± 18 K. Interestingly, green data-model residuals between the $D_1$ and $D_2$ lines are similarly structured, thus imperfections in the model's fitting cannot be attributed simply to spectral noise. This phenomenon will be investigated in detail in Section 6 and discussed in Section 7.

Figure 6 shows the distribution of effective temperatures for sodium gas around Mercury's disk, as obtained through the above described approach. Seeing far above the horizon varied from 0.91-1.1″ night to night. It is dubious to apply these values at Mercury's airmass >7, but this expands the fiber footprint by at least 600 km full width half maximum. Seeing blurs the spatial resolution on- and off-disk, but the Doppler velocity at points tangent to the surface is less affected due to limb-brightening. Near the subsolar point and footprints of the magnetic cusps, linewidth derived effective temperatures are in good agreement with the MESSENGER scale height derived temperature of ~1200 K (Cassidy et al., 2015), with no high temperature component detected. $D_2/D_1$ ratios below the expected value of 1.6 imply line cores can exceed unity optical depth in these regions, however this agreement suggests that distortion of the line shape does not significantly bias the effective temperature retrieval. Toward the poles and nightside limb, effective temperatures increase to ~1400 K. These locations probe the pronounced exospheric sodium enhancement at cold pole longitudes (dashed blue) and dawn local times identified by Cassidy et al. (2016) using MESSENGER limb scans. Offsets of 1 $R_M$ (2440 km) from each pole were intended to sample the hot scale height Na population that Vervack et al. (2010) reported during the MESSENGER M3 flyby. Above 800 km altitudes, these authors reported that the e-folding atmospheric falloff increased to ~500 km, more than twice the scale height they found near the surface. Modeling by Suzuki et al. (2020) shows that this increase is best fit by a secondary population liberated via micrometeoroid impact vaporization, or possibly solar wind sputtering. Curiously, an analogous effect is not borne in the Doppler broadening data, which show a modest increase of just 4% at high altitudes above Mercury's poles. A dramatic increase in Doppler broadening was observed down the northern lobe of the exotail, however, a characteristic that will be examined further in the subsequent section.

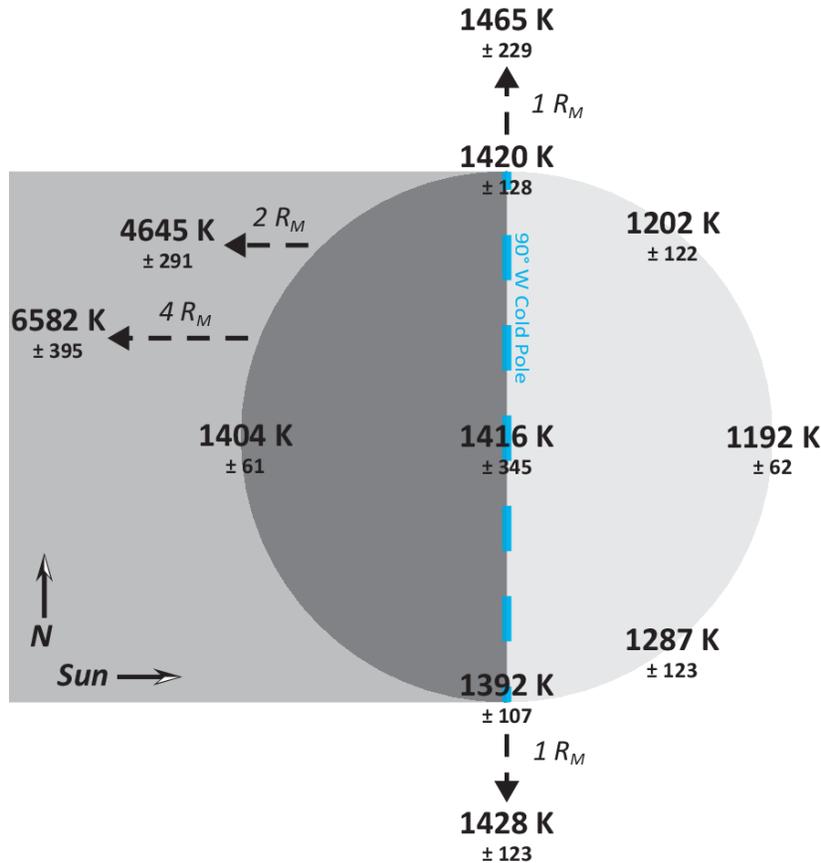

*Figure 6.* Sodium effective temperatures measured by KPF on 22-23 March 2024 with observer phase angle 85-90° and TAA 27-34°. In this geometry, the 90° W cold pole lies directly at the dawnside terminator. Quantities are the average of the individually fit $D_1$ and $D_2$ results, with some regions also being the average of multiple measurements. Sodium effective temperatures near ~1200 K are coldest at the subsolar point, increasing 200-300 K with latitude and on the nightside. Effective temperatures increase sharply down the tail.

Despite R~100,000 resolving power and the high sensitivity of Keck's large aperture, the KPF spectra were found to be insufficient for useful constraints on potassium Doppler broadening. Lierle et al. (2022) showed that potassium is the least energetic metal in Mercury's exosphere. Their Doppler broadening measurements of the K D lines placed potassium in near thermal equilibrium with Mercury's surface and not measurably wider than the R~150,000 instrumental resolution of EXPRES. Thus, while KPF's inability to quantify potassium energies is disappointing, it is not surprising.

### 6. Sodium & Potassium Doppler Broadening in the Exotail.

Two observing campaigns were conducted at the LDT using the Extreme Precision Spectrometer, one 2022 April 25-28 and the other 2023 April 9-12. Spectra were acquired at greatest elongation in both campaigns. In 2022, the observer phase angle ranged from 88-99° through TAA 65-80°. In 2023, the observer phase angle ranged from 89-103° through TAA 50-66. Figure 7 shows the regions around each of the four studied emission lines for a 30 s integration of the disk center. Analysis of EXPRES spectra follows KPF methodology outlined in Section 4. Exospheric emissions are isolated by fitting and subtracting off a model solar reflectance spectrum (orange). $D_2$ is the brighter transition for both sodium and potassium, but in this observing geometry K $D_2$ is partially absorbed by telluric $O_2$ at 7663.73Å. Telluric absorptions are corrected for using SELENITE, a model developed and tested specifically for EXPRES data (Leet et al., 2019). Pointing downtail, the scattered solar continuum quickly falls off. Integration times ranged from 10 s on disk to 300 s downtail.

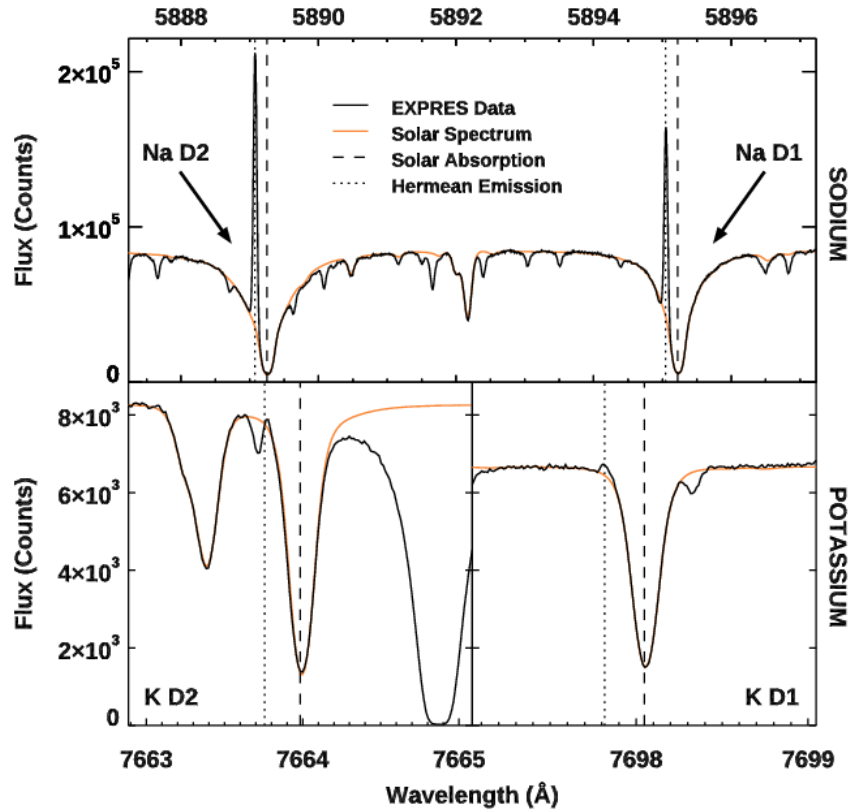

*Figure 7.* Extracted 30 s LDT/EXPRES spectrum acquired 10 April 2023 on Mercury's dayside. Solar continuum is removed using the orange fit, and subsequent corrections for telluric $H_2O$ (around sodium) and $O_2$ (around potassium) are treated independently within the EXPRES data pipeline (Leet et al., 2019).

Figure 8 shows pointing for all 10 April 2023 EXPRES observations. Mercury's disk is 7.3″ in diameter and the triangular EXPRES fiber housing is shown as it appears in the fast tip-tilt guide camera, but note that the hexagonal fiber itself is a smaller 0.9″ (600 km) in diameter. The spectrum in Fig. 7 (black) was acquired above mid-morning on the equatorial dayside, with five subsequent spectra sampling out to 4.3 $R_M$ down the southern lobe of the tail. The blurred image of Mercury in Figure 8 is an aligned co-addition of all the fast tip-tilt guider frames taken over the course of this evening. Pointing moved tailward as image quality deteriorated with Mercury setting near the horizon.

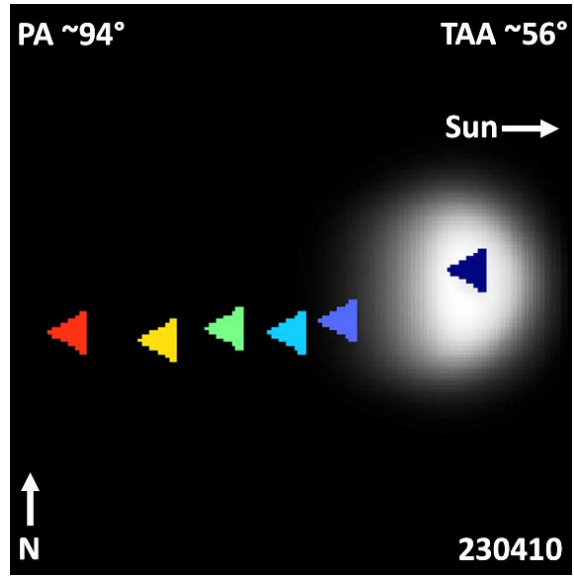

*Figure 8.* Co-added image of the EXPRES fast tip-tilt guide mirror, showing pointing locations sampling up to 4.3 $R_M$ down the southern lobe of the tail on 10 April 2023. Triangles indicate the 0.9" fiber housing.

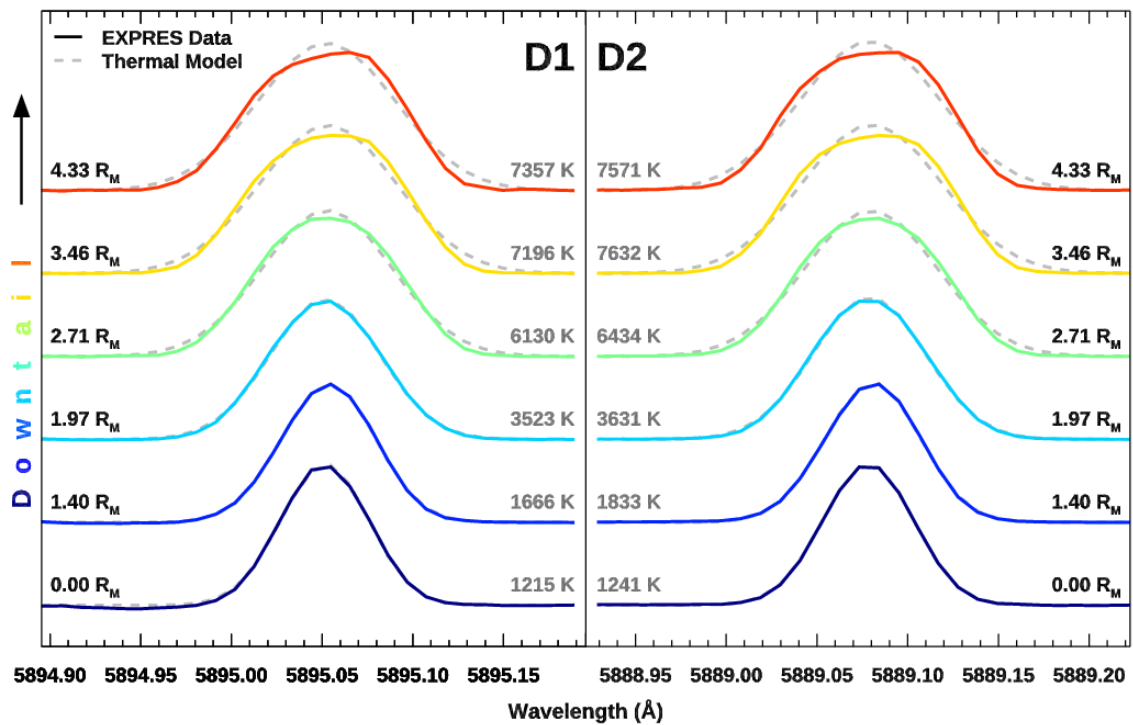

*Figure 9. EXPRES line profiles of sodium D1 and D2 at increasing downtail distance from Mercury. Spectra are color-coded with their pointing in Figure 8. Data are overlaid with thermal forward models in grey, illustrating how, as line profiles broaden, they also become increasingly nonthermal in shape. Best-fit temperatures are given in grey. Downtail distances are zeroed at the center of the planetary disk. Note that these line profiles are normalized in amplitude so that they may be easily compared; actual radiance drops with distance downtail.*

Figure 9 shows isolated sodium D line profiles as a function of antisunward distance in the exotail. On-disk, the sodium effective temperature is found to be 1228 ± 48 K, in close agreement with MESSENGER scale heights (Cassidy et al., 2015) and the low and mid-latitude weighted average of 1233 ± 14 K obtained via KPF. Linewidths increase dramatically downtail, where both $D_1$ and $D_2$ exceed effective temperatures of 7000 K. Coverage of this morphology with downtail distance is more comprehensive than obtained via KPF, but note the similar observing geometry and effective temperatures. $D_2$ appears slightly hotter than $D_1$ for each pointing. While $D_2$ saturates at a lower column abundance than $D_1$, extremely rarified columns downtail preclude such an opacity effect. Rather, their discrepancy may be a systematic consequence of the hyperfine strengths and wavelengths applied, or any subtle variation in the instrumental LSF between the two D lines.

Grey overlaid forward models of the line shape in Figure 9 highlight growing disagreement between the measurements and a theoretical Doppler broadened profile. Measured profiles farthest downtail exhibit depleted cores, and moving away from line center, there is an over intensity followed by an under intensity in the line wings. This structure mimics that seen in the green KPF residuals in Figure 5, but becomes quite prominent at greater distance, where line profiles become less Gaussian and more boxcar shaped. An additional redward skew develops past 3 $R_M$, indicating that gas velocities in the tail are not symmetric along the line of sight, despite the near 90° phase angle. This is evidence that sodium gas was predominantly redshifted in the tail, which cannot be explained by an exosphere that is axisymmetric with respect to the Sun-Mercury vector.

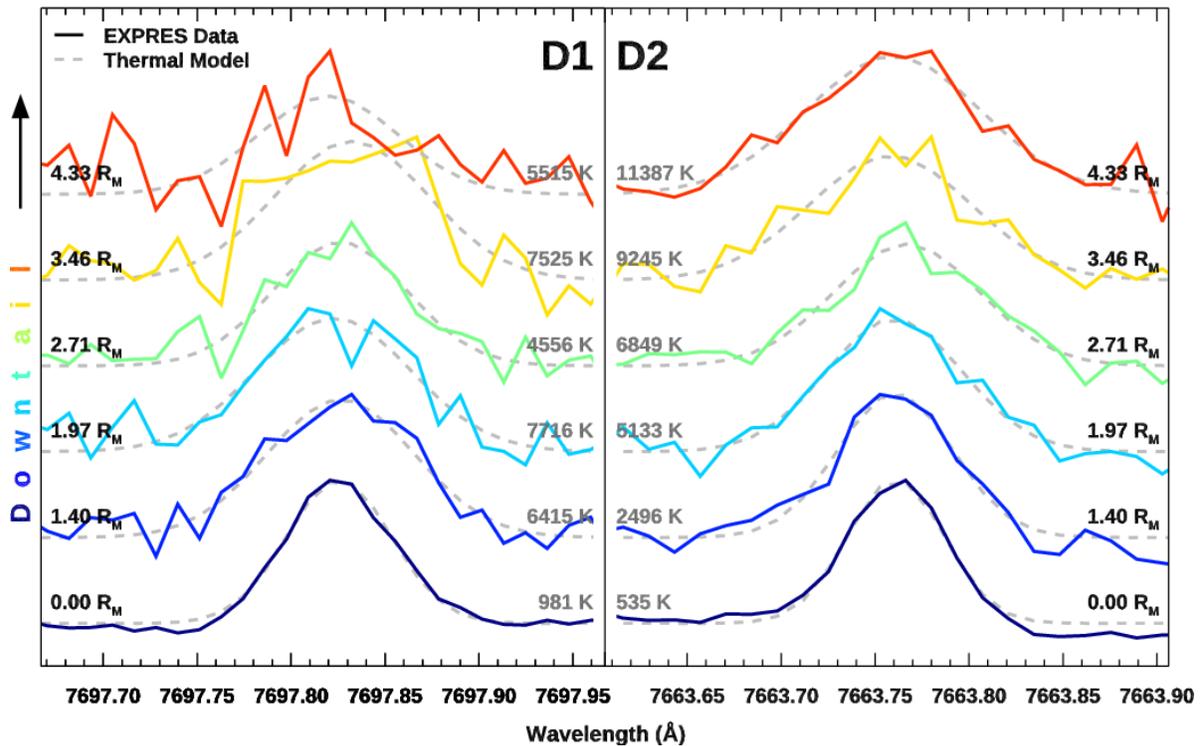

***Figure 10.*** *EXPRES line profiles of potassium D1 and D2 at increasing downtail distance from Mercury. As with sodium, potassium linewidths broaden with downtail distance, most evident in the brighter D2 line.*

Contemporaneous potassium line profiles are shown in Figure 10. These emissions are fainter than sodium (the Na/K column ratio derived in Section 3 is ~95 in the tail) and so their line profiles are much noisier. Still, linewidth growth with downtail distance is clearly evident. On disk, the doublet-averaged potassium effective temperature is 758 ± 585 K, consistent with the cold linewidths measured by Lierle et al. (2022) and not much wider than the instrumental limiting resolution. Far downtail, $D_1$ approaches the continuum noise floor, but brighter $D_2$ emission increases to >10,000 K, an even steeper climb than exhibited in sodium. While $D_2$ remains bright enough for the model to converge at these distances, its emission offers insufficient signal to noise for any analogous conclusions to be made on the thermality of the line shapes. Unlike sodium, one of the potassium D lines does not appear systematically warmer, at least within what can be discerned given their high noise levels.

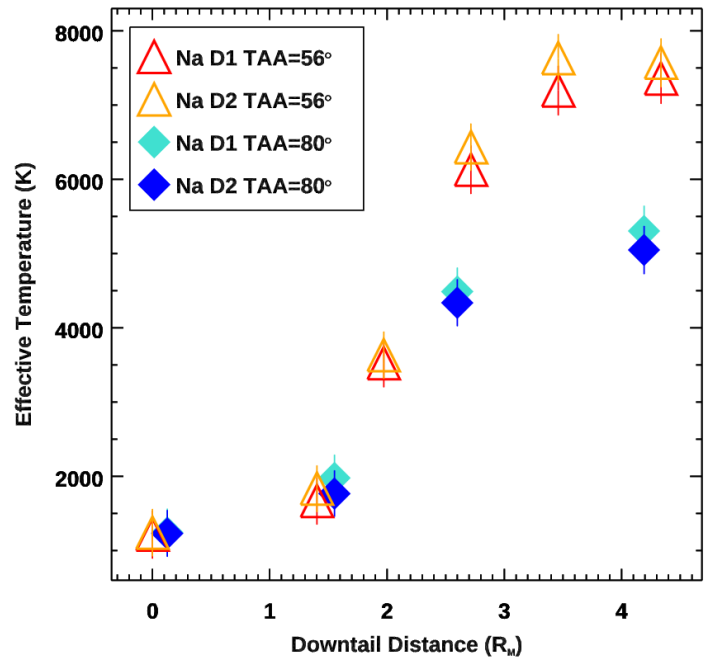

*Figure 11. Sodium effective temperatures as a function of downtail distance for two true anomaly angles. Effective temperature levels off around 3.5 $R_M$ downtail for both, though this level-off effective temperature is higher for TAA = 56° than for TAA = 80°.*

Plotting sodium effective temperature against downtail distance in Figure 11 reveals that the Doppler broadening along Mercury's tail is variable. 10 April 2023 observations discussed so far in this section and 28 April 2022 observations had similar planetary geometry aside from their difference in true anomaly angle of 56° and 80°, respectively. Both of these nights exhibit non-linear growth in their linewidths, where effective temperatures rapidly increase downtail before reaching a horizontal asymptote just past 3 $R_M$. At TAA = 56°, this leveling off occurs at ~7500 K, while for TAA = 80° it occurs lower, at ~5000 K.

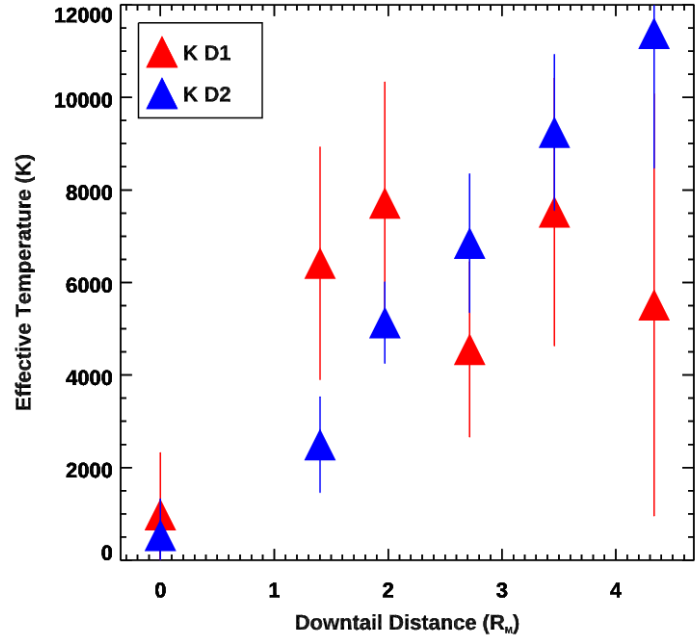

*Figure 12. Potassium effective temperatures from EXPRES as a function of downtail distance on 2023 April 10 at TAA = 56°. Data from 28 April 2022 were omitted since the lesser photon scatting at TAA = 80° reduced potassium escape and radiance in the tail.*

Due to the lower brightness, it is difficult to discern whether potassium follows non-linear behavior with distance that is similar to sodium. Radiation pressure on potassium peaks at 48° TAA and while this forcing at 80° TAA is only 11% less than that at 56° TAA, the difference appears critical for K escape. The K tail was only detected very near peak radiation pressure in the long-slit campaign, and while K emissions are discernable the exotail in 2022 spectra, their line profile has weak signal to noise, precluding any reliable effective temperature estimation. Hence, only 2023 measurements are plotted in Figure 12. The brighter $D_2$ line appears to linearly increase in effective temperature with downtail distance, while $D_1$ lacks a clear functional form. With this ambiguity, it is possible that potassium linewidth growth levels off similarly to sodium downtail, but any such change falls within the 1σ uncertainty on these effective temperatures.

## 7. Interpretation and Discussion

### 7.1 Sources Supplying Na and K to the Exosphere

Sodium and potassium effective temperatures derived by high resolution spectroscopy are most consistent with supply to the exosphere by photon-stimulated desorption (PSD). The low to mid-latitude weighted average of 1233 ± 14 K for sodium and 758 ± 585 K for potassium are both consistent with laboratory time-of-flight measurements by Yakshinskiy & Madey (2003) (cf. Lierle et al., 2022 Figure 11). Their experiments ejected alkali atoms ionically bound to oxygen in silica quartz. The alkalis were neutralized by electron transfer from Auger decay, which results in a repulsive state between neighboring atoms and thus liberation, the same underlying physical mechanism as in PSD. The 1.66 ratio in observed Na/K thermal velocities is similar to their experimental value of 1.54, and nearly the 1.30 ratio of their inverse root of atomic mass, as expected if alkali atoms were imparted equal energies. Na and K bonding sites are expected to be ionic on the topmost monolayers in regolith where photons can reach, regardless of whether the atoms are chemically bound in mineral form or adsorbed from the gas phase exosphere (Madey et al., 1998). The low threshold energies for desorption of alkali atoms correspond to far blue wavelengths where solar irradiance is constant, and so their PSD into Mercury's exosphere is not expected to vary significantly with solar activity.

A photodesorbed supply of sodium must be reconciled with the strong brightness enhancements near Mercury's magnetic cusps measured by ground-based observers. Since ion sputtering is too energetic, only three alternatives exist. First, electron stimulated desorption could be more efficient than previously thought, since this releases alkali atoms with the same energy as PSD while also concentrated at the cusps. Second, these enhancements could be produced by solar wind ions 'gardening' the regolith, increasing solid state diffusion from depth and enhancing supply to the topmost monolayers where atoms desorb. Third, the sodium enhancements near the cusps may not relate to a magnetospheric influence at all, but rather to sodium surface heterogeneity. Where measurements exist in the northern hemisphere, Na is concentrated at the pole (Peplowski et al., 2014), while solar photon flux is concentrated at the equator. The balance between these where PSD is maximal may well occur at mid latitudes that are near the magnetic cusps only by coincidence. Dynamic short-term changes in the exosphere reported by many observers (Leblanc et al., 2009; Mangano et al., 2013; Massetti et al., 2017; Orsini et al., 2018) favors some magnetospheric influence, as in the first two scenarios. The lack of episodic changes in sodium from the MESSENGER orbiter (Cassidy et al., 2021) and the static nature of exospheric enhancements seen in ground-based measurements (Schmidt et al., 2020) favors the third. It is worth noting that the tangent points of the MESSENGER MASCS measurements were concentrated at low latitudes, making the detection of solar wind and magnetospheric influence difficult. Still, the cause of these brightness enhancements in sodium will likely be left an open question, at least until Mercury Sodium Atmospheric Spectral Imager (MSASI; Yoshikawa et al., 2010) orbital measurements from BepiColombo.

Near the day-night terminator, Na effective temperatures increase by ~200 K, but remain well below energetic ejection processes like micrometeoroid vaporization or sputtering. At 1 $R_M$ above Mercury's poles, well beyond the 800 km altitudes where Vervack et al. (2010) began to observe a hot sodium component, Doppler broadened effective temperatures average only 1462 ± 87 K. While this effective temperature could be achieved in the transition region between a bound cold population and extended hot gas, line profiles here begin to exhibit a flattened line core (e.g. residuals in Figure 5) that becomes more pronounced downtail. A superposition of sources as a convolution of two Gaussian components would not produce the observed line core deficiency.

Instead, the higher temperatures in these regions are interpreted as an effect of the loss of the least energetic atoms during transport from the dayside by strong solar radiation pressure. Low energy atoms experience more numerous surface encounters during transport. MESSENGER analysis found no evidence for a thermalized component (Cassidy et al., 2015) and models suggest that any accommodation towards the local surface temperature during bouncing is weak (Burger et al., 2010; Mouawad et al., 2011; Schmidt, 2013). While atoms encountering Mercury's hot dayside surface are expected to bounce, nearly 20% of Na atoms adsorb to smooth 500 K surfaces, with roughness only adding to this (Yakshinskiy & Madey, 2005). Surface adsorption in this way, effectively a form of gravitational filter, alters the gas velocity distribution during transport across the planet; cold atoms are more likely to adsorb, while hotter atoms travel farther between bounces.

This may explain the measured increase in effective temperature with distance from the subsolar point. Na winds towards the nightside could, in principle, offer a measurable bulk Doppler shift for such a scenario. Evidence for this is mixed (Leblanc et al., 2013; Potter et al., 2009), but the observation is challenging because measuring a radial component necessitates small solar elongation angles. Additionally, it is unclear if past measurements accounted for the ~0.6 km/s gravitational redshift described in Section 2, since the shift from the solar well is commonly used as a reference for the rest frame emission wavelength.

7.2 Effective Temperatures in the Escaping Exotail

Both sodium and potassium gases exhibit dramatic Doppler broadening between the dayside and 4.3 $R_M$ down Mercury's exotail. Initially, we considered this as a potential signature of heating by photon recoils, which impart a nearly isotropic momentum transfer on the gas population. As atoms are accelerated down the tail, they Doppler shift out of the Fraunhofer absorption wells, increasing their photon scattering rates and thus potential heating. The effective temperature would then level off when the gas population is Doppler shifted into the solar continuum and positive feedback ceases. However, even during maximum radiation acceleration Na atoms would not Doppler shift to scatter solar continuum photons until >80 $R_M$ downtail (cf. Schmidt et al., 2010 Figure 6). Strong disagreement with the observed 3.5 $R_M$ level-off point in Figure 11 effectively rules out recoil heating as a viable explanation.

Observed Doppler broadening downtail is instead theorized to result again from the gravitational filtering of low energy sodium and potassium atoms introduced in Section 7.1, now at high altitudes above the surface. If a 1200 K, loosely Maxwellian population of sodium gas is produced at the surface, the coldest of those atoms will quickly re-impact and be adsorbed, while hotter atoms will reach higher altitudes, the most energetic escaping. When EXPRES is pointed at Mercury's surface near the subsolar point, the entire source velocity distribution is contained within its line of sight. When it is pointed above the limb, or down the tail, only the fraction of the original 1200 K population with energies sufficient reach that altitude is observed. This concept is illustrated in Figure 13, but note that its portrayal is overly simplistic in some aspects. First, a fixed velocity distribution is shown at each altitude, while in reality the distribution shifts towards lower velocities balancing the gained gravitational potential energy at altitude. Second, the velocity distribution will also evolve in time from radiation acceleration, which is challenging to treat analytically. Incorporation of these coupled effects prompts detailed modeling, and so Figure 13 is intended merely to convey the most general mechanics.

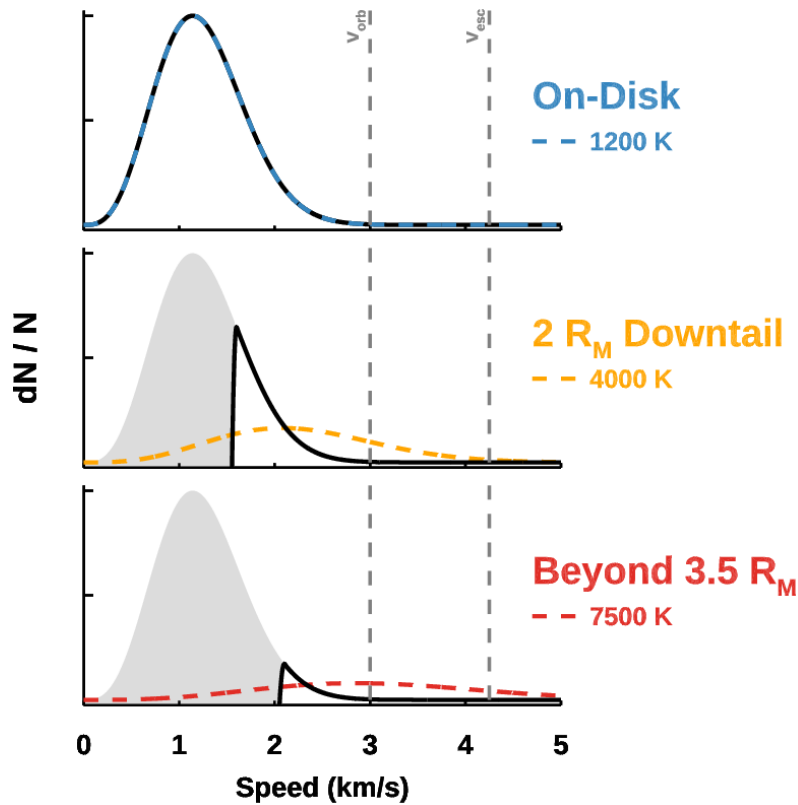

*Figure 13. Illustration of the proposed effect of gravitational filtering of particle speeds with downtail distance. The shaded region indicates the fraction of the original ~1200 K source population that is gravitationally bound below that altitude. Distribution of sodium atoms measured at altitude is represented as the black curve for three pointing locations. Maxwell-Boltzmann flux distributions at the measured effective temperatures for each location are represented by dashed curves. The discrepancy between these Maxwellians and the proposed population after gravitational filtering also illustrates why line shapes become nonthermal down the tail. $v_{esc}$ corresponds to the sodium escape speed, and $v_{orb}$ is the speed required to enter a satellite trajectory, from which atoms can escape with the aid of radiation acceleration.*

When pointing far enough above the planet's surface, only atoms with energies high enough to escape Mercury's gravity are observed. This causes the leveling off of sodium effective temperature in Figure 11 at around 3.5 $R_M$. The farthest extent of gravitationally bound sodium, then, is placed at < 3.5 $R_M$, beyond which essentially all atoms escape. Radiation acceleration strongly aids escape, so this apex altitude would depend on location and be most compressed at the sub-solar point. Seasonal variation is expected as well, since radiation acceleration varies by an order of magnitude over the course of a Mercury year. This could qualitatively explain the different leveling-off effective temperatures between TAAs 56° and 80° in Fig. 11.

Gravitational filtering produces non-Maxwellian energy distributions at altitude, portrayed in Figure 13, even if the surface supply velocity distribution is Maxwellian. This effect is evident in the Na line profiles in Figure 9. Above the planet's limb, the line core is flattened as low energy atoms are filtered out, while the high energy atoms populating the wings remain. This gives the illusion of a broader, nonthermal emission line despite no heating occurring. Nonthermality is most evident beyond 3.5 $R_M$, where the escaping gas is close to monoenergetic and the measured line profiles begin to resemble a boxcar convolved with the instrumental LSF.

Recent analysis of PHEBUS data from the BepiColombo flybys found that if emission in the c404 channel were attributed to potassium, the temperature corresponding to its measured altitude profile is 1500 K (Robidel et al., 2023). As a scale height derived temperature, this measurement is not expected to be affected by the observational broadening discussed herein, since the gravitational filtering effect is inherent in their applied Chamberlain model. Rather, a plausible explanation stems from the PHEBUS c404 channel passing lines from several species besides K due to its broad filter width. Robidel et al. (2023) conclude that, given higher than expected radiance for K, the profile measured by PHEBUS may be blend of K with Mn, and possibly Al or $Ca^+$. Each of these species would be more energetic than K, alleviating some tension between their 1500 K value, which is exclusive to a potassium emission source, and the $758 \pm 585$ K derived here. Additionally, BepiColombo's closest approach during the flybys analyzed was on Mercury's nightside. Our sodium line profiles show greater Doppler broadening near the day-night terminator relative to other regions of the dayside, and so by analogy it is conceivable that potassium may also be locally warmer in regions the flybys sampled.

7.3 Doppler Shifts and Anisotropic Sources

The redward skew of the EXPRES sodium profiles in Figure 9 indicates a brighter population of atoms traveling from dawn to dusk across the tail than vice versa. This is not an effect of the bulk downtail atomic motion, since at phase angle 94° the tiny projection of this motion along the line of sight would appear blueshifted. However, the sub-observer longitude was 95° W, meaning the 90° W cold pole had just crossed the dawn terminator into sunlight. According to Cassidy et al. (2016) and ground-based work by Milillo et al. (2021), cold pole longitudes at 90° and 270° W produce a pronounced dawn enhancement after they rotate into sunlight near 30° TAA. Monte Carlo models show that sodium trajectories that are pulled in the shadow by Mercury's gravity and those that escape traverse to the opposite hemisphere before being forced down the tail upon reencountering sunlight and radiation pressure (Schmidt, 2013; Smyth & Marconi, 1995). In this way the measured line asymmetry in sodium is plausibly an indication of an enhanced dawnside source of Na atoms associated with the cold-pole longitudes.

## 8. Conclusions

Mercury's proximity to the Sun makes it an excellent environment for observing the effect of gravitational filtering on emission line profiles. This phenomenon should occur in all planetary exospheres, surface-bounded or otherwise, and so the capability to observe evolving line profiles with altitude offers a powerful diagnostic for atmospheric escape.

In this work, a combination of long-slit spectroscopy and precision point measurements of resolved line profiles was used to examine the distribution and linewidth of sodium and potassium emissions in Mercury's exosphere and exotail. Results are summarized as:

1) Mercury's potassium exotail is detected for the first time, both in high resolution point measurements up to ~4 $R_M$ antisunward, and in long slit spectroscopy up to ~10 $R_M$ antisunward.

2) The weighted average of sodium gas effective temperatures at low and mid-latitudes is 1233 ± 14 K along the noon meridian, in good agreement with MESSENGER-based scale heights here.

3) The weighted average of sodium gas effective temperatures at the dawnside terminator and poles is 1406 ± 14 K, substantially hotter than measurements elsewhere on the dayside. We propose this difference arises from gas transport toward the terminator by strong radiation pressure, and the loss of the least energetic atoms to Mercury's surface during this transport.

4) Sodium gas at high altitudes 1 $R_M$ above Mercury's poles is hotter still, with an effective temperature of 1451 ± 71 K. However, this effective temperature via Doppler broadening is well below the expectation from past measurements of extended scale heights here.

5) Effective temperatures of alkali gases obtained via Doppler broadening increase dramatically between Mercury's dayside and its escaping tail. Averaging results of each doublet, from the dayside to 4.3 $R_M$ downtail sodium increases from 1228 K to 7464 K, and potassium increases from 758 K to 8451 K.

6) Line profiles of Na D in the exotail appear distinctly non-Maxwellian (*cf.* Fig. 9). Their shape changes from a Gaussian to boxcar with increasing distance.

7) Results 3-6 above are interpreted as an effect of gravitational filtering of the velocity distribution function, in which low energy atoms fall to Mercury's surface and high energy atoms escape. The observations at a given altitude reflect what velocity distribution remains after Mercury's gravity removes the least energetic population component.

8) The sodium effective gas temperature becomes invariant between 3.5 and 4.3 $R_M$ down the exotail. This is interpreted as the distance at which the gas population is entirely escaping and gravitational filtering is no longer altering the velocity distribution function. The ballistic apex of bound particle trajectories in this season is hence < 3.5 $R_M$.

9) Morphology of line profiles when transitioning from gas that is predominantly bound to gas that is predominantly escaping has not been previously reported observationally. Alkali D lines at Mercury offer a "textbook" example of a universal phenomenon that should be broadly prevalent in planetary exospheres, but has not been described previously. The use of Doppler broadening to probe a non-thermal velocity distribution function effectively constitutes a new technique for characterizing atmospheric escape.

Future work will attempt to recreate the observed line profiles via numerical Monte Carlo modeling of Mercury's sodium and potassium exospheres. This method can explicitly track the evolution of the speed distribution, including the influence of radiation acceleration and source angular distribution. Results will help constrain the effects of radiation pressure on emission line morphology and provide insight into the seasonal variability of this phenomenon.

**Acknowledgments**

We thank the KPF instrument team, particularly Howard Isaacson and Jack Wright, as well as staff astronomer Josh Walawender, for welcoming the challenge of observing Mercury. We also thank members of the EXPRES instrument team, particularly Debra Fischer, John Michael Brewer, Lily Zhao, Ryan Blackman, Allen Davis, Ryan Petersburg, and Andrew Szymkowiak, for their support of the observations at LDT and assistance with data analysis. PL acknowledges support from the Massachusetts Space Grant and NASA/NExSCI Keck Award under JPL contract #1707133. CS gratefully acknowledges support from NASA under awards 80NSSC22K1303, 80NSSC21K1019 and 80NSSC21K0051.

Some of the data presented in this paper were obtained at the W.M. Keck Observatory, which is operated as a scientific partnership among the California Institute of Technology, the University of California, and the National Aeronautics and Space Administration. Observing time for this project was allocated by all three institutions, in part thanks to the Twilight Zone Program. NASA Keck time is administered by the NASA Exoplanet Science Institute. The Observatory was made possible by the generous financial support of the W. M. Keck Foundation. The authors wish to recognize and acknowledge the very significant cultural role and reverence that the summit of Maunakea has always had within the indigenous Hawaiian community. We are most fortunate to have the opportunity to conduct observations from this mountain.

**Open Research**

The KPF data analyzed within this work are available on the W. M. Keck Observatory Archive (koa.ipac.caltech.edu). All other data analyzed within this work, including analysis codes in IDL, are publicly archived at Lierle (2025).

accepted for publication in *Journal of Geophysical Research: Planets*

**References**

Bida, T. A., Killen[2], R. M., & Morgan[3], T. H. (2000). *Discovery of calcium in Mercury's atmosphere*. NATURE (Vol. 404). Retrieved from www.nature.com

Brown, R. A., & Yung, Y. L. (1976). *Io, Its Atmosphere and OPTICAL EMISSIONS*. Tucson: University of Arizona Press.

Burger, M. H., Killen, R. M., Vervack, R. J., Bradley, E. T., McClintock, W. E., Sarantos, M., et al. (2010). Monte Carlo modeling of sodium in Mercury's exosphere during the first two MESSENGER flybys. *Icar*, *209*(1), 63–74. https://doi.org/10.1016/J.ICARUS.2010.05.007

Burger, M. H., Killen, R. M., McClintock, W. E., Merkel, A. W., Vervack, R. J., Cassidy, T. A., et al. (2014). Seasonal variations in Mercury's dayside calcium exosphere. *Icar*, *238*, 51–58. https://doi.org/10.1016/J.ICARUS.2014.04.049

Cassidy, T. A., Schmidt, C. A., Merkel, A. W., Jasinski, J. M., & Burger, M. H. (2021). Detection of Large Exospheric Enhancements at Mercury due to Meteoroid Impacts. *PSJ*, *2*(5), 175. https://doi.org/10.3847/PSJ/AC1A19

Cassidy, Timothy A., Merkel, A. W., Burger, M. H., Sarantos, M., Killen, R. M., McClintock, W. E., & Vervack, R. J. (2015). Mercury's seasonal sodium exosphere: MESSENGER orbital observations. *Icarus*, *248*. https://doi.org/10.1016/j.icarus.2014.10.037

Cassidy, Timothy A., McClintock, W. E., Killen, R. M., Sarantos, M., Merkel, A. W., Vervack, R. J., & Burger, M. H. (2016). A cold-pole enhancement in Mercury's sodium exosphere. *Geophysical Research Letters*, *43*(21). https://doi.org/10.1002/2016GL071071

Chamberlain, J. W. (1963). *PLANETARY CORONAE AND ATMOSPHERIC EVAPORATION\**. Pktw. Space Sci (Vol. 11).

Coddington, O. M., Richard, E. C., Harber, D., Pilewskie, P., Woods, T. N., Chance, K., et al. (2021). The TSIS-1 Hybrid Solar Reference Spectrum. *GeoRL*, *48*(12), e91709. https://doi.org/10.1029/2020GL091709

Emerson, & D. (1996). Interpreting Astronomical Spectra. *Ias*, 472. Retrieved from https://ui.adsabs.harvard.edu/abs/1996ias..book.....E/abstract

Falke, S., Tiemann, E., Lisdat, C., Schnatz, H., & Grosche, G. (2006). Transition frequencies of the D lines of K39, K40, and K41 measured with a femtosecond laser frequency comb. *Physical Review A - Atomic, Molecular, and Optical Physics*, *74*(3). https://doi.org/10.1103/PhysRevA.74.032503

Gibson, S. R., Howard, A. W., Rider, K., Roy, A., Edelstein, J., Kassis, M., et al. (2020). Keck Planet Finder: design updates. *Https://Doi.Org/10.1117/12.2561783*, *11447*, 848–865. https://doi.org/10.1117/12.2561783

Killen, R. M., Potter, A., Fitzsimmons, A., & Morgan, T. H. (1999). Sodium D2 line profiles: clues to the temperature structure of Mercury's exosphere. *Planetary and Space Science*, *47*(12), 1449–1458. https://doi.org/10.1016/S0032-0633(99)00071-9

Killen, R. M., Bida, T. A., & Morgan, T. H. (2005). The calcium exosphere of Mercury. *Icarus*, *173*(2), 300–311. https://doi.org/10.1016/j.icarus.2004.08.022

Killen, R. M., Killen, & M., R. (2016). Pathways for energization of Ca in Mercury's exosphere. *Icar*, *268*, 32–36. https://doi.org/10.1016/J.ICARUS.2015.12.035

Killen, R. M., Morrissey, L. S., Burger, M. H., Vervack, R. J., Tucker, O. J., & Savin, D. W. (2022). The Influence of Surface Binding Energy on Sputtering in Models of the Sodium Exosphere of Mercury. *The Planetary Science Journal*, *3*(6), 139. https://doi.org/10.3847/PSJ/AC67DE

Leblanc, F., Doressoundiram, A., Schneider, N., Massetti, S., Wedlund, M., Ariste, A. L., et al. (2009). Short-term variations of Mercury's Na exosphere observed with very high spectral resolution. *GeoRL*, *36*(7), L07201. https://doi.org/10.1029/2009GL038089

Leblanc, F., Schmidt, C., Mangano, V., Mura, A., Cremonese, G., Raines, J. M., et al. (2022). Comparative Na and K Mercury and Moon Exospheres. *Space Science Reviews 2022 218:1*, *218*(1), 1–56. https://doi.org/10.1007/S11214-022-00871-W

Leblanc, Francois, Chaufray, J. Y., Doressoundiram, A., Berthelier, J. J., Mangano, V., López-Ariste, A., & Borin, P. (2013). Mercury exosphere. III: Energetic characterization of its sodium component. *Icar*, *223*(2), 963–974. https://doi.org/10.1016/J.ICARUS.2012.08.025

Leet, C., Fischer, D. A., & Valenti, J. A. (2019). Toward a Self-calibrating, Empirical, Light-weight Model for Tellurics in High-resolution Spectra. *The Astronomical Journal*, *157*(5), 187. https://doi.org/10.3847/1538-3881/AB0D86

Lierle, P. (2025). Mercury Exosphere Linewidths Data 2017-2023 and Codes [Data set]. *Zenodo*. https://doi.org/10.5281/ZENODO.14728010

Lierle, P., Schmidt, C., Baumgardner, J., Moore, L., Bida, T., & Swindle, R. (2022). The Spatial Distribution and Temperature of Mercury's Potassium Exosphere. *The Planetary Science Journal, Volume 3, Issue 4, Id.87, <NUMPAGES>9</NUMPAGES> Pp.*, *3*(4), 87. https://doi.org/10.3847/PSJ/AC5C4D

Lierle, P., Schmidt, C., Baumgardner, J., Moore, L., & Lovett, E. (2023). Rapid Imaging Planetary Spectrograph. *Publications of the Astronomical Society of the Pacific*, *135*(1051). https://doi.org/10.1088/1538-3873/acec9f

Madey, T. E., Yakshinskiy, B. V., Ageev, V. N., & Johnson, R. E. (1998). Desorption of alkali atoms and ions from oxide surfaces: Relevance to origins of Na and K in atmospheres of Mercury and the Moon. *JGR*, *103*(E3), 5873–5888. https://doi.org/10.1029/98JE00230

Mangano, V., Massetti, S., Milillo, A., Mura, A., Orsini, S., & Leblanc, F. (2013). Dynamical evolution of sodium anisotropies in the exosphere of Mercury. *P&SS*, *82*, 1–10. https://doi.org/10.1016/J.PSS.2013.03.002

Mangano, V., Massetti, S., Milillo, A., Plainaki, C., Orsini, S., Rispoli, R., & Leblanc, F. (2015). THEMIS Na exosphere observations of Mercury and their correlation with in-situ magnetic field measurements by MESSENGER. *Planetary and Space Science*, *115*. https://doi.org/10.1016/j.pss.2015.04.001

Massetti, S., Mangano, V., Milillo, A., Mura, A., Orsini, S., & Plainaki, C. (2017). Short-term observations of double-peaked Na emission from Mercury's exosphere. *GeoRL*, *44*(7), 2970–2977. https://doi.org/10.1002/2017GL073090

McClintock, W. E., & Lankton, M. R. (2007). The mercury atmospheric and surface composition spectrometer for the MESSENGER mission. *Space Science Reviews*, *131*(1–4), 481–521. https://doi.org/10.1007/S11214-007-9264-5/METRICS

Milillo, A., Mangano, V., Massetti, S., Mura, A., Plainaki, C., Alberti, T., et al. (2021). Exospheric Na distributions along the Mercury orbit with the THEMIS telescope. *Icarus*, *355*. https://doi.org/10.1016/j.icarus.2020.114179

Morton, D. C. (2003). Atomic Data for Resonance Absorption Lines. III. Wavelengths Longward of the Lyman Limit for the Elements Hydrogen to Gallium. *ApJS*, *149*(1), 205–238. https://doi.org/10.1086/377639

Mouawad, N., Burger, M. H., Killen, R. M., Potter, A. E., McClintock, W. E., Vervack, R. J., et al. (2011). Constraints on Mercury's Na exosphere: Combined MESSENGER and ground-based data. *Icar*, *211*(1), 21–36. https://doi.org/10.1016/J.ICARUS.2010.10.019

Orsini, S., Mangano, V., Milillo, A., Plainaki, C., Mura, A., Raines, J. M., et al. (2018). Mercury sodium exospheric emission as a proxy for solar perturbations transit. *NatSR*, *8*(1), 928. https://doi.org/10.1038/S41598-018-19163-X

Orsini, S., Mangano, V., Milillo, A., Mura, A., Aronica, A., De Angelis, E., et al. (2024). Remote sensing of mercury sodium exospheric patterns in relation to particle precipitation and interplanetary magnetic field. *NatSR*, *14*(1), 30728. https://doi.org/10.1038/S41598-024-79022-W

Peplowski, P. N., Evans, L. G., Stockstill-Cahill, K. R., Lawrence, D. J., Goldsten, J. O., McCoy, T. J., et al. (2014). Enhanced sodium abundance in Mercury's north polar region revealed by the MESSENGER Gamma-Ray Spectrometer. *Icarus*, *228*. https://doi.org/10.1016/j.icarus.2013.09.007

Potter, A. E., Killen, R. M., & Morgan, T. H. (2007). Solar radiation acceleration effects on Mercury sodium emission. *Icarus*, *186*(2), 571–580. https://doi.org/10.1016/j.icarus.2006.09.025

Potter, A. E., Killen, R. M., Potter, A. E., & Killen, R. M. (2008). Observations of the sodium tail of Mercury. *Icar*, *194*(1), 1–12. https://doi.org/10.1016/J.ICARUS.2007.09.023

Potter, A. E., Morgan, T. H., & Killen, R. M. (2009). Sodium winds on Mercury. *Icarus*, *204*(2), 355–367. https://doi.org/10.1016/j.icarus.2009.06.028

Quémerais, E., Chaufray, J. Y., Koutroumpa, D., Leblanc, F., Reberac, A., Lustrement, B., et al. (2020). PHEBUS on Bepi-Colombo: Post-launch Update and Instrument Performance. *Space Science Reviews*. https://doi.org/10.1007/s11214-020-00695-6

Robidel, R., Quémerais, E., Chaufray, J. Y., Koutroumpa, D., Leblanc, F., Reberac, A., et al. (2023). Mercury's Exosphere as Seen by BepiColombo/PHEBUS Visible Channels During the First Two Flybys. *Journal of Geophysical Research: Planets*, *128*(12), e2023JE007808. https://doi.org/10.1029/2023JE007808

Schmidt, C. A. (2013). Monte Carlo modeling of north-south asymmetries in Mercury's sodium exosphere. *JGRA*, *118*(7), 4564–4571. https://doi.org/10.1002/JGRA.50396

Schmidt, C. A. (2022). Doppler-Shifted Alkali D Absorption as Indirect Evidence for Exomoons. *Frontiers in Astronomy and Space Sciences*, *9*, 801873. https://doi.org/10.3389/fspas.2022.801873

Schmidt, C. A., Wilson, J. K., Baumgardner, J., & Mendillo, M. (2010). Orbital effects on Mercury's escaping sodium exosphere. *Icar*, *207*(1), 9–16. https://doi.org/10.1016/J.ICARUS.2009.10.017

Schmidt, C. A., Baumgardner, J., Mendillo, M., & Wilson, J. K. (2012). Escape rates and variability constraints for high-energy sodium sources at Mercury. *Journal of Geophysical Research: Space Physics*, *117*(3). https://doi.org/10.1029/2011JA017217

Schmidt, C. A., Baumgardner, J., Moore, L., Bida, T. A., Swindle, R., & Lierle, P. (2020). The Rapid Imaging Planetary Spectrograph: Observations of Mercury's Sodium Exosphere in Twilight. *The Planetary Science Journal*, *1*(1). https://doi.org/10.3847/psj/ab76c9

Smyth, W. H., & Marconi, M. L. (1995). Theoretical Overview and Modeling of the Sodium and Potassium Atmospheres of Mercury. *ApJ*, *441*, 839. https://doi.org/10.1086/175407

Suzuki, Y., Yoshioka, K., Murakami, G., & Yoshikawa, I. (2020). Seasonal Variability of Mercury's Sodium Exosphere Deduced From MESSENGER Data and Numerical Simulation. *Journal of Geophysical Research: Planets*, *125*(9). https://doi.org/10.1029/2020JE006472

Vervack, R. J., McClintock, W. E., Killen, R. M., Sprague, A. L., Anderson, B. J., Burger, M. H., et al. (2010). Mercury's Complex Exosphere: Results from MESSENGER's Third Flyby. *Sci*, *329*(5992), 672. https://doi.org/10.1126/SCIENCE.1188572

Yakshinskiy, Boris V., & Madey, T. E. (2005). Temperature-dependent DIET of alkalis from SiO2 films: Comparison with a lunar sample. In *Surface Science* (Vol. 593, pp. 202–209). https://doi.org/10.1016/j.susc.2005.06.062

Yakshinskiy, B. V., & Madey, T. E. (2003). DIET of alkali atoms from mineral surfaces. In *Surface Science* (Vol. 528, pp. 54–59). https://doi.org/10.1016/S0039-6028(02)02610-9

Yoshikawa, I., Korablev, O., Kameda, S., Rees, D., Nozawa, H., Okano, S., et al. (2010). The Mercury Sodium Atmosphere Spectral Imager (MSASI) onboard BepiColombo/MMO. *Cosp*, *38*(1–2), 6. https://doi.org/10.1016/J.PSS.2008.07.008